\documentclass[useAMS,usenatbib]{mn2e}
\usepackage{amssymb}
\usepackage{amsfonts}
\usepackage{graphicx}

\def\vec#1{{\mbox{\boldmath$#1$}}}

\title[Cosmography with cluster strong lensing]{Cosmography
with cluster strong lensing}

\author[Gilmore \& Natarajan]{James Gilmore$^1$\thanks{E-mail: james.gilmore@yale.edu} and Priyamvada
Natarajan$^{1,2,3}$\thanks{E-mail: priya@astro.yale.edu}\\
$^1$Department of Physics, Yale University, P.O. Box
208120, New Haven, CT 06520-8120, USA\\
$^2$Department of Astronomy, Yale University, P.O.
Box 208101, New Haven, CT 06520-8101, USA \\
$^3$Radcliffe Institute for Advanced Study, 10 Garden Street,
Cambridge, MA 02138}
\begin{document}

\date{\today}
\pagerange{\pageref{firstpage}--\pageref{lastpage}} \pubyear{2009}
\maketitle \label{firstpage}

\begin{abstract}
By stacking an ensemble of strong lensing clusters, we demonstrate
the feasibility of placing constraints on the dark energy equation
of state. This is achieved by using multiple images of sources at
two or more distinct redshift planes. The sample of smooth clusters
in our simulations is based on observations of massive clusters and
the distribution of background galaxies is constructed using the
Hubble Deep Field. Our source distribution reproduces the observed
redshift distribution of multiply imaged sources in Abell 1689. The
cosmology recovery depends on the number of image families with
known spectroscopic redshifts and the number of stacked clusters.
Our simulations suggest that constraints comparable to those derived
from other competing established techniques on a constant dark
energy equation of state can be obtained using 10 to 40 clusters
with 5 or more families of multiple images. We have also studied the
observational errors in the image redshifts and positions. We find
that spectroscopic redshifts and high resolution {\it Hubble Space
Telescope} images are required to eliminate confidence contour
relaxation relative to the ideal case in our simulations. This
suggests that the dark energy equation of state, and other
cosmological parameters, can be constrained with existing {\it
Hubble Space Telescope} images of lensing clusters coupled with
dedicated ground-based arc spectroscopy.
\end{abstract}


\section{INTRODUCTION}

Current efforts in observational cosmology are directed towards
characterizing the energy content of the Universe. The discovery of
the accelerating expansion of the Universe inferred from the Hubble
diagram for SN type Ia
\citep{riess98,perlmutter99,tonry2003,riess04,riess07} combined with
constraints from Cosmic Microwave Background Radiation from the
Wilkinson Microwave Anisotropy Probe
\citep{WMAPcosm2003,WMAPcosm2006,Hinshaw09}, from cosmic shear
observations
\citep{bacon00,kaiser00,vanwaerbeke00,wittman00,semboloni05},
cluster baryon fractions \citep{allen04}, galaxy surveys
\citep{efstat2002,seljak05} suggests that $\sim 70 $ per cent of the
energy in the Universe is in the form of dark energy. Dark energy is
best described by the equation of state that relates its pressure,
$P$, to its energy density, $\rho$, via $P = \textrm{w}_\textrm{x}
\rho$, with $\textrm{w}_\textrm{x} = -1$ corresponding to the case
of the cosmological constant.

Massive foreground clusters often produce many sets of highly
distorted arcs and multiple images of background galaxies
\citep{bland92}. The lensing effect is determined by the mass
distribution, the precise configuration of the lens and source with
respect to the observer, and the angular diameter distances between
the lens, source and observer. When at least two sources at distinct
redshift are strongly lensed in a cluster, the relative positions of
the arcs depend on the ratios of angular diameter distances. In
principle therefore, cosmological parameters can be constrained from
the lensing configuration.

Constraining the geometry and matter content of the Universe using
multiple sets of arcs in cluster lenses has been explored by
\citet{paczy1981,link1998,coor1999,gols2002,sere2002,souc2004}.
Recently, \citet{beis2006} has suggested using multiple images
produced by galaxy lenses instead of clusters and attempted an
application to the case of the lensing system HST\,14176+5226.
\citet{link1998} showed that the cosmological sensitivity of the
angular size-redshift relation could be exploited using sources at
distinct redshifts and developed a methodology to simultaneously
invert the lens and derive cosmological constraints.
\citet{sere2002} applied the technique to the cluster Cl\,0024+1654
and found support for an accelerating Universe. \citet{gols2002}
showed that the recovery of cosmological parameters was feasible
with 3 sets of multiple images for a single simulated cluster.
\citet{souc2004} then applied the technique to the lensing cluster
Abell 2218 and using 4 multiple image systems at distinct redshifts,
find $\Omega_{\rm M}<0.37$ and $w<-0.80$ assuming a flat Universe.
Recently, a double Einstein ring has been discovered and applied to
cosmography \citep{Gavazzi08}. As the system has only two sources,
only weak cosmography constraints were obtained.

There have been several studies that have explored the
cross-sections of multiple image formation using simulated clusters
from cosmological \textit{N}-body simulations.
\citet{mene2005a,mene2005b} and \citet{macc2005} studied the
efficiency of numerical clusters to produce multiple images in
different dark energy models. They found the expected abundance of
arcs with a given length-to-width ratio and the lensing
cross-sections depend on the equation of state of dark energy. These
statistical studies of multiple image formation concluded that
strong lensing offers a plausible method to discriminate amongst
various dark energy models.

In this paper we study two basic aspects of the non-statistical
approach to cosmography with cluster strong lensing (CSL), which
have yet to be addressed. These are: 1) How many clusters must be
individually marginalized, with the resulting $\chi^2$ from each
cluster added together to form a final $\chi^2$ for cosmology
recovery (i.e. stacked), in the ideal case to obtain cosmological
constraints comparable with other methods and 2) The required
accuracy of the observational parameters, i.e. the source redshift
and image positions. By studying these two aspects separately from
the subtleties involved with cluster mass distribution
reconstruction, we can make definitive statements on the type of
cluster surveys required for a first application of CSL.

We do not model the lensing effects induced by the distribution of
substructure along the line of sight. This additional complication
has been examined in earlier work by \citet{wbo2005} and
\citet{daha2005}, who both used ray tracing methods through $N$-body
simulations. Simulating a representative shell of the Universe
centered at $z = 1$, \citet{wbo2005} found that approximately
10\%-15\% of the lensing mass was to be found in substructure along
the line of sight, for about one third of lensing configurations
with a source redshift above $z_s=2.5$. In particular, it was shown
that substructure along the line of sight is limited to a finite
number of lens planes, typically one or two additional planes, when
line of sight substructure played a significant role in lensing.
They have shown explicitly that $<20\%$ of the contribution to the
convergence arises from secondary lens planes. For cluster strong
lensing the critical factor is not the error induced in the
convergence but rather the perturbations caused in the positions of
the multiple images due to the presence of matter in additional lens
planes. A detailed and exhaustive set of cosmological simulations
that will enable the matter distribution of a typical line of sight
to be averaged over many representative lines of sight is needed to
accurately quantify this effect. In \citet{daha2005} it was shown
that systematic errors in the cosmology recovery could result, if
the line of sight substructure was not taken into account.

The outline of this paper is as follows: we begin with an
introduction to the basics of strong lensing and dark energy in
Section~2. Then in Section~3 we discuss image families, and define
the cosmological family ratio, which CSL constrains. The methodology
used to generate smooth simulated clusters is described in Section~4
and Section~5, along with the $\chi^2$ minimization scheme used in
the cosmology recovery. We present the results of recovering the
input cosmology in Section~6, and study the dependence of the
recovery on lens parameters in Section~7. We then study the allowed
errors in the source redshift and image positions in Section~8. This
is followed by a discussion of the simplifying assumptions made in
this work in Section~9. We conclude with a summary of our technique,
and the implications for future survey strategies.

\section{STRONG GRAVITATIONAL LENSING AND DARK ENERGY}

The gravitational lens equation is a mapping from the source plane
at redshift $z_{\rm s}$ to the image plane (or lens plane) at
redshift $z_{\rm l}$ and is given by,
\begin{equation}\label{lensequation}
\vec{\theta}\,=\,\vec{\beta}\,+\,\vec{{\alpha}}(\vec{\theta},\xi;M)
\end{equation}
where $\vec{\theta}$ is the angular position of the image,
$\vec{\beta}$ is the unobserved source position, $\vec{\alpha}$ is
the deflection angle which depends on $\xi$ the reduced angular
diameter distance, $M$ the mass distribution of the lens and
$\vec{\theta}$. The cosmological dependence is contained in $\xi$,
which we call the reduced angular diameter defined by:
$\xi\,=\,D(0,z_{\rm l})D(z_{\rm l},z_{\rm s})/D(0,z_{\rm s})
\,\equiv\,D_{\rm ol}\,D_{\rm ls}/D_{\rm os}$, where $D(z_1,z_2)$ is
the angular diameter distance between redshift $z_1$ and $z_2$, and
$D_{\rm ol}$, $D_{\rm ls}$ and $D_{\rm os}$ are the observer-lens,
lens-source and observer-source angular diameter distances
respectively. In a flat homogeneous Friedmann-Robertson-Walker
cosmology the angular diameter distance is given by,
\begin{equation}\label{angulardiameter}
    D(z_1,z_2)=\frac{c H_0^{-1}}{1+z_2}\int_{z_1}^{z_2}\!\!\!dz
\left(\Omega_{\rm M}(1+\!z)^{3}+\Omega_{\rm
X}(1+\!z)^{3(\textrm{w}_{\rm X}+1)}\right)^{-1/2}
\end{equation}
where $c$ is the speed of light; $H_0\,=\,100\,h\,{\rm km\,s^{-1}}$
is the present day value of the Hubble constant, and $\Omega_{\rm
M}$ and $\Omega_{\rm X}$ are the fractional contributions of the
total matter and dark energy respectively, in units of the critical
density and $\textrm{w}_{\rm X}$ is the dark energy equation of
state. Note that we do not consider evolving dark energy models in our
simulations, since there is currently no observational evidence for
such a time evolution in the dark energy equation of state
\citep{sahni06,Komatsu2009}.

To further explore the dependence of the lens equation on
cosmological parameters, we note that the deflection angle can be
written in terms of the gradient of the reduced gravitational
potential $\varphi_M(\vec{\theta},\xi)$:
$\vec{\nabla}\varphi_M(\vec{\theta},\xi)\,=\,\vec{{\alpha}}(\vec{\theta},\xi;M)$.
The reduced gravitational potential can be related to the projected
surface mass density along the line of sight,
$\Sigma_M(\vec{\theta})$, through the Poisson equation, where the
projected gravitational potential $\phi_{M}(\vec{\theta})$ is
defined by
\begin{equation}\label{projgravpot}
    \varphi_M(\vec{\theta},\xi)=\xi\,\frac{2}{c^2}
    \,\phi_M(\vec{\theta}).
\end{equation}
In the thin lens approximation the projected gravitational potential
is a solution of: $\vec{\nabla}^2\varphi_M(\vec{\theta},\xi)=8\pi
G\,\xi\,\Sigma_M(\vec{\theta})/c^2$. The total deflection
$(\vec{\theta}-\vec{\beta})$ is therefore determined both by the
cosmological parameters (via $D_{\rm ol}D_{\rm ls}/D_{\rm os}$) and
the projected surface mass density of the intervening lens. For a
given configuration of the source, lens and observer, the critical
value of the surface mass density is:
\begin{equation}\label{eqn:sigmacrit}
\Sigma_{\rm crit}\,=\,\frac{c^2}{4\,\pi\,G}\,\frac{1}{\xi}.
\end{equation}
The region where the projected surface mass density of a lens
exceeds this critical value defines the strong lensing regime.
Background sources that are aligned behind this region are likely to
be multiply imaged, magnified and distorted.

There are hence two fundamental equations which define the
gravitational lensing dependence on the angular diameter distances
$D_{\rm ol}, D_{\rm ls}, D_{\rm os}$. The first is the lens mapping,
equation~(\ref{lensequation}), which defines the configuration of
the images produced by a lensing cluster, and the second is
equation~(\ref{eqn:sigmacrit}) which defines the strong lensing
regime. The cosmological dependence of strong lensing is discussed
further in Section~\ref{sec:cosmofamilyratio}.

\section{THE COSMOLOGICAL FAMILY RATIO}\label{sec:cosmofamilyratio}

\begin{figure*}
\includegraphics[scale=1]{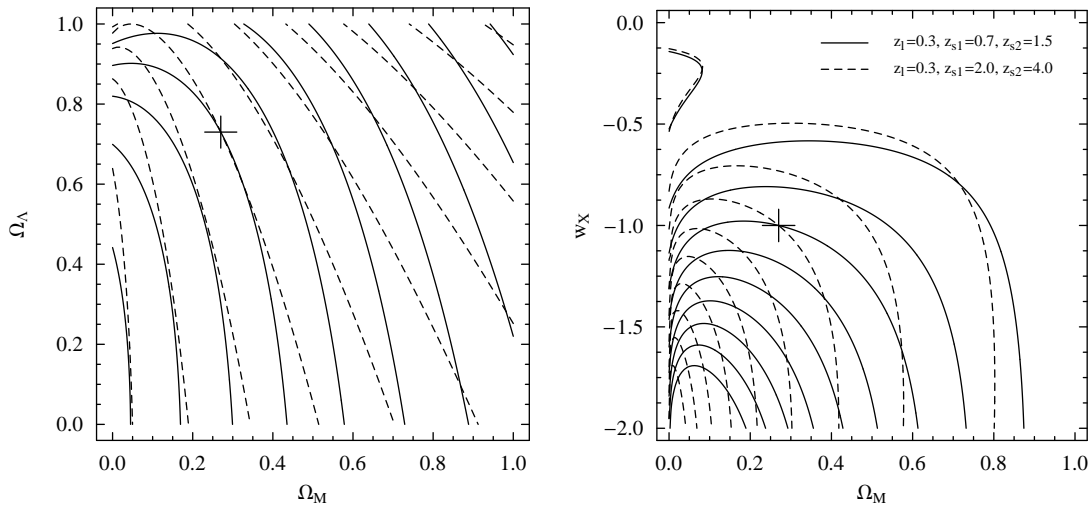}
\caption{Contours of the cosmological family ratio $\Xi(z_{\rm
l},z_{\rm s1},z_{\rm s2};\Omega_{\rm M},\Omega_{\rm
X},\textrm{w}_{\rm X})$ on the $\Omega_{\rm M}$-$\Omega_{\Lambda}$
and $\Omega_{\rm M}$-$\textrm{w}_{\rm X}$ planes. The cross marks
the input cosmology. The redshift of the lensing cluster is fixed at
$z_{\rm l} = 0.3$. Solid contours correspond to sources at $z_{\rm
s1} = 0.7$, $z_{\rm s2} = 1.5$ and dashed to sources at $z_{\rm
s1}=2.0$, $z_{\rm s2}=4.0$ in both panels.} \label{fig:familyratio}
\end{figure*}

\subsection{Multiple image families}

Consider a lensing configuration where at least two sources have
been strongly lensed by a cluster. Once the multiple images of each
source have been identified, we can write the source and the
associated image positions with a family index $f$. The lens
equation is then,
\begin{equation}\label{lenseqintrofinot}
    \vec{\beta}_{f} = \vec{\theta}_{f,i} - \vec{\nabla}
\varphi_M(\vec{\theta}_{f,i},\xi),
\end{equation}
where $\vec{\theta}_{f,i}$ denotes the position of the $i^{\rm th}$
image of family $f$. When there are $k$ multiple images detected in
family $f$, we can sum over all the images in
equation~(\ref{lenseqintrofinot}). Substituting into
equation~(\ref{projgravpot}) for the reduced gravitational potential
allows the summed equation to be written in terms of the line of
sight projected gravitational potential as follows:
\begin{equation}\label{lenscos1}
\sum^{k}_{i=1}\vec{\theta}_{f,i}=\frac{2}{c^2}\left(\frac{D_{\rm
ls}}{D_{\rm ol}D_{\rm os}}\right)
\sum^{k}_{i=1}\vec{\nabla}\phi_M(\vec{\theta}_{f,i})+k\,\vec{\beta}_{f}.
\end{equation}
Note that equation~(\ref{lenscos1}) is a non-linear function of the
image positions, since the sum of the image positions depends on the
projected cluster potential gradient.

Given a set of input cosmological parameters, the lens redshift and
a source redshift for a single family of images, i.e. $(z_{\rm
l},z_{\rm s},\Omega_{\rm M},\Omega_{\rm X},\textrm{w}_{\rm X})$, the
angular diameter distance ratio is specified and constant. The
cluster potential is conventionally normalized by the square of the
velocity dispersion, which we denote by $v$. In
equation~(\ref{lenscos1}) the reduced angular diameter distance
$\xi$ and the cluster mass normalization $v^2$ are degenerate when
recovering cosmological parameters. \textit{Therefore, cosmological
parameters cannot be disentangled from the cluster mass
normalization with only one family of multiple images.}  Possible
solutions to remove this degeneracy are considered by
\citet{gols2002}. They show an additional prior on the mass profile
is required and this must come from observations independent of
strong lensing. For parametric models, this is typically a total
mass determination from the detected X-ray emission of the cluster.

\subsection{Two families of multiple images}

Now consider the case when two strongly lensed sources are observed
as multiple image families in the same cluster lens at two
redshifts. Let family 1 have a total of $m$ observed images and
family 2 have $n$ observed images. We require that at least two
images from each are observed, i.e. $m,n\ge2$.  Taking the ratio of
family 1 to family 2 (equation [\ref{lenscos1}]) we obtain,
\begin{equation}\label{lenscos2}
\left\{\frac{D_{\rm ls1}}{D_{\rm os1}}\frac{D_{\rm os2}}{D_{\rm
ls2}}\right\} \frac
{\sum^{m}_{i=1}\vec{\nabla}\phi_M(\vec{\theta}_{1,i})}
{\sum^{n}_{j=1}\vec{\nabla}\phi_M(\vec{\theta}_{2,j})}= \frac
{-m\vec{\beta}_{1}+\sum^{m}_{i=1}\vec{\theta}_{1,i}}
{-n\vec{\beta}_{2}+\sum^{n}_{j=1}\vec{\theta}_{2,j}}.
\end{equation}
Note that we now have a ratio of two factors of
$\vec{\nabla}\phi_M(\vec{\theta})$. This clearly means that the
degeneracy between the cosmological parameters and the cluster
potential normalization is broken when 2 families of multiple images
are observed in a cluster lens.

In equation~(\ref{lenscos2}), the term in braces is a function of
the angular diameter distances only. This quantity is fundamental to
the determination of cosmological parameters through CSL and we
define this term as the family ratio and denote it by $\Xi$:
\begin{equation}\label{xidefn}
    \Xi(z_{\rm l},z_{\rm s1},z_{\rm s2};\Omega_{\rm M},\Omega_{\rm X},\textrm{w}_{\rm X})=\frac{D(z_{\rm l},z_{\rm
s1})}{D(0,z_{\rm s1})}\frac{D(0,z_{\rm s2})}{D(z_{\rm l},z_{\rm
s2})}
\end{equation}
where $z_{\rm l}$ is the lens redshift, $z_{\rm s1}$ and $z_{\rm
s2}$ are the two source redshifts, and $D(z_{1},z_{2})$ is the
angular diameter distance defined in
equation~(\ref{angulardiameter}). The cosmological family ratio can
be written in terms of the reduced angular diameter distance,
$\Xi\,=\,\xi_1/\xi_2$, where the subscripts 1 and 2 here denote the
family index.

We plot $\Xi$ in Fig.~\ref{fig:familyratio} for a fixed lens
redshift of $z_{\rm l}=0.3$ and differing source redshifts. Since
the non-evolving dark energy component with $\textrm{w}_{\rm
X}\,=\,-1$ can also be interpreted as the cosmological constant
$\Omega_{\Lambda}$, we plot the contours of the cosmological family
ratio in both the $\Omega_{\rm M}$-$\Omega_{\Lambda}$ plane in
addition to the $\Omega_{\rm M}$-$\textrm{w}_{\rm X}$ plane. Note
that the contours in the $\Omega_{\rm M}$-$\Omega_{\Lambda}$ plane
are not as sensitive to the lens-source configuration as they are in
the $\Omega_{\rm M}$-$\textrm{w}_{\rm X}$ plane.

\section{SIMULATIONS OF STRONG LENSING CLUSTERS}\label{simulations}

As shown in the previous section, the cosmological family ratio,
$\Xi$, depends on the source redshifts and the lens redshift (see
Fig.~\ref{fig:familyratio}). It is therefore clear that the
distribution of sources in magnitude and redshift as well as the
cluster redshift distribution are key inputs in our simulations.
Therefore we base our simulations of strong lensing clusters on high
resolution observational data from \textit{HST} images. We note that
although the cluster normalization is irrelevant to our method for
constraining the cosmological parameters, we still construct this
quantity rigorously. This ensures the simulations produce lensing
clusters with cross-sections similar to those of observed clusters.

\subsection{Source magnitude and redshift
distribution}\label{sec:sourcedistconstruct}

In order to simulate the lensed images produced by a massive
cluster, it is necessary to characterize the number density, $N$, of
the sources behind the lens. We consider background point sources
for which the number density is a function of magnitude and
redshift, i.e. $N\,=\,N(m,z)$. Furthermore, space-based positional
accuracy is assumed throughout this paper. We must therefore draw
redshifts and magnitudes from a space-based source distribution. The
Hubble Deep Field (HDF) \cite{hdf1996} and HDF-South (HDF-S) WFPC2
\cite{hdfs2000} are the ideal data sets to derive the source
distribution. It is also possible to use the recent COSMOS data to
derive the redshift and magnitude source distributions, see
\citet{Gabasch08}. In particular we have checked that our source
redshift distribution, which is derived below, is equivalent to that
of \citet{Gabasch08}.

We employ the WFPC2 HDF and HDF-S photometric redshift catalogues
derived by the Stony Brook collaboration \citep{fern1999,yahata2000}
and characterize the HDF sources by a distribution function. This
method is chosen since we wish to analyse an ensemble of clusters,
where each cluster lenses a unique source configuration.
Characterizing the source distribution by a function is preferable
in our case, since we do not want to bias the results by lensing
large scale structures such as voids or filaments. Note that there
is evidence of a weak cluster in the HDF \citep{vill1997}.

To model the source distribution from the HDF and HDF-S data, we
first make two cuts to the catalogues. Initially, we drop all
objects with magnitudes greater than $AB(8140) = 28.00$. Next we
select objects in the redshift interval $0.4 \leq z \leq 5.0$. The
lower bound is chosen because the lensing efficiency is negligible
for low source redshifts. The upper redshift bound is chosen because
the photometric redshift determination is unreliable for $z\gtrsim
5.0$ and most observed multiply imaged sources behind massive
clusters have redshift $z \lesssim 5.0$. This yields $1688$ objects
when the catalogues are merged. The corresponding projected source
number density is $215\,$arcmin$^{-2}$ for the ensemble catalogue.

We now bin the merged catalogue in magnitude from $AB(8140)\,=\,19.0
- 28.0$ in intervals of $0.5$ and in redshift from $z\,=\,0.4 - 5.0$
in intervals of $0.2$. These data are then fit with functions of the
form,
\begin{eqnarray}\label{eqn:HDFfit}
\nonumber N(m,z)\,&=&\,f\,\exp[-\,a\,(m-b)^{s}]\,\exp[-\,c\,(z-d)^{t}] \\
  &&\times\exp[-\,g\,(m-b)^{u}(z-d)^{v}],
\end{eqnarray}
where the exponents in the exponential terms, $(s, t, u, v)$, can
take an even integer value. The constants $a, b, c, d,$ and $f$ are
determined in the fitting procedure. We simply use the lowest
$\chi^2$ fit to construct a probability distribution function in the
simulations. This is $(s, t, u, v)=(4,2,8,2)$, with the other
parameters given by: $a=1.607\times10^{-4}$, $b=30.70$,
$c=9.97\times10^{-2}$, $d=0.507$, $g=2.197\times10^{-7}$ and
$f=18.462$.

Notably, the $\chi^2$ of the above fit is approximately one half of
the $\chi^2$ that is obtained when the binned magnitude and redshift
data are decoupled and fitted on an individual basis, i.e. $N(z)$
and $N(m)$. It is therefore clear that to avoid a systematic bias in
the source generation, a joint fit in magnitude and redshift space
is required.

\subsubsection{Comparison to Abell 1689}

\begin{figure*}
\includegraphics[scale=1]{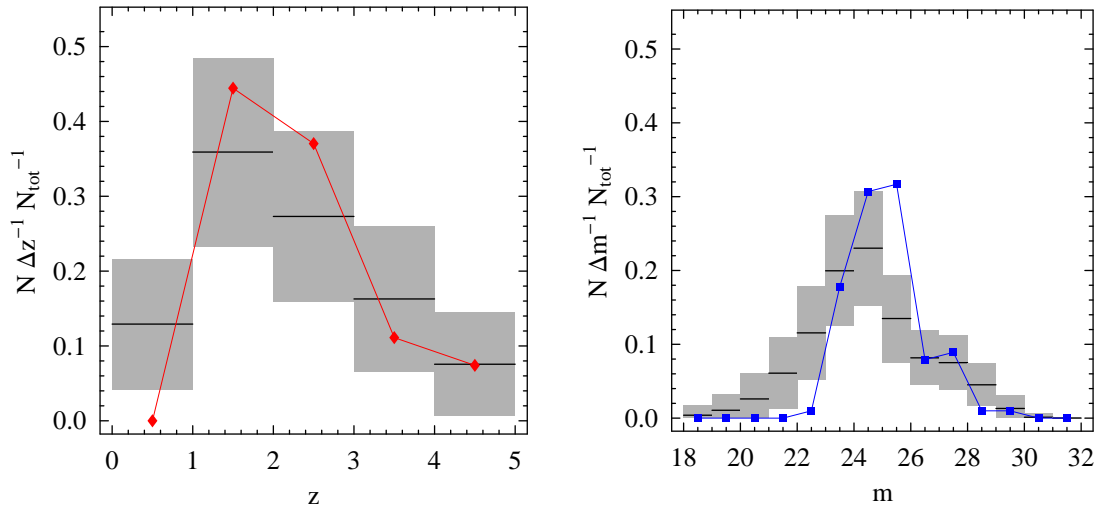}
\caption{Comparison of image properties using the lensed HDF derived
distribution and the Abell 1689 data of \citet{broad2005}. In each
bin, the simulated data has a mean given by the solid horizontal
line and the gray rectangle represent $\pm$1$\sigma$. {\it Left
panel:} Redshift distributions. The Abell 1689 data are plotted as
solid diamonds. {\it Right panel:} Magnitude distributions. The
solid squares are the Abell 1689 data.} \label{figAbell}
\end{figure*}

Deep ACS imaging of Abell 1689 has allowed identification of more
than 30 families of multiply images systems
\citep{broad2005,limo2007}. It is clear that Abell 1689 is the ideal
candidate to compare with our source distribution. We base our
analysis on \citet{broad2005}, however our conclusions are unchanged
if we use the analysis of \citet{limo2007}. With the lens model as a
guide \citet{broad2005} have identified multiple images to
$I\,=\,29.82$. This means images with magnitudes comparable to and
less than the magnitude limit of our HDF catalogues,
$AB(8140)\,=\,28.00$, are detected in the ACS observations. This
means we can compare our source distribution to the properties of
Abell 1689 directly.

In order to compare the source characteristics of Abell 1689 with
our source distribution, we need to lens this distribution through a
mass model. \citet{broad2005} have published an Navarro, Frenk, \&
White (NFW) fit (see Section~\ref{sec:NFW}) to approximate the true
mass profile. Using a radially averaged (and therefore circular)
mass profile, they provide the tangential and radial critical
curves, computed at redshift $z\,=\,3.0$. They also determine the
NFW scale radius $r_{\rm s}\,=\,473^{+197}_{-169}\,$kpc and we adopt
$r_{\rm s}\,=\,473\,$kpc. We then adjust an NFW fit until the two
critical curve locations are reproduced.

Since we are using a smooth profile, the properties of Abell 1689 we
compare our source distribution with cannot depend on the small
scale details of the mass distribution. There are two quantities
which satisfy this requirement: 1) the total number of lensed
sources and 2) their redshift distribution. A test of the HDF
constructed source distribution would be the ability to reproduce
these two quantities for simulated lenses similar to Abell 1689.
Note that we do expect some systematic effects due to the lack of
assumed substructure in the modeling. In particular, the fact that
Abell 1689 is non-circular, and is known to have images which are
influenced by substructure, means any comparison between the actual
data and our model (which is circular, and has no structure) will
exhibit residuals.

To compare our simulations with the Abell 1689 sources, we take
$z_{\rm best}$ as the source redshift for each image from table~2 of
\citet{broad2005}. Also, because the HDF catalogues used to derive
the source distribution have been cut at $z\,=\,5.0$, we drop all
sources with redshift $z\,>\,5.0$. The Abell 1689 source redshifts
are then binned in intervals of $\Delta z\,=\,1.0$ from $z\,=\,0.0$
to $z\,=\,5.0$. For proper comparison with the simulations, the bin
counts are normalized by the total number of families in Abell 1689.

We simulate the smooth Abell 1689 NFW profile for 500 source
configurations generated with our source distribution. We then lens
the sources independently for each configuration, using the
procedure explained in Section~\ref{sec:clusterandimage}. Multiply
imaged sources are identified and binned identically as the Abell 1689
sources.

The average number of multiply imaged sources in the 500 simulations
is $N_{\rm sim}\,=\,16 \pm 4$ per simulation. The minimum and
maximum number of families are 7 and 28 respectively and of the 500
configurations 102 exhibit 20 or more families of images. To
$z\,=\,5.0$ there are 27 families in the \citet{broad2005} analysis.
Although the mean number for our generated images is lower than that
of the 27 strongly lensed sources in Abell 1689 by $\sim\,3\sigma$,
we do not consider this a serious problem. The total number of
strongly lensed sources is within factor of 2 in 69 per cent of the
500 realizations.

We now examine the redshift distribution of the lensed sources in
the simulations. The configurations are combined to produce a
distribution of redshift probabilities in each bin. This allows the
mean and standard deviation to be computed, see Fig.~\ref{figAbell}.
Our simulated redshift distribution lies within 1$\sigma$ of the
Abell 1689 sources, except for the $0\le z<1$ bin for which there
are no observed multiply imaged sources in Abell 1689.

Finally we discuss the image properties themselves. Since the mass
model does not include substructure, we do not expect to reproduce
the source multiplicity distribution or the image magnitude
distribution of Abell 1689 accurately. For a circular NFW mass profile,
lensed sources with multiplicity of at most 3 can only be produced,
whereas higher multiplicities are observed in Abell 1689. However, as
shown in Fig.~\ref{figAbell}, the image magnitude distribution
as generated through our source distribution does match that of
Abell 1689, and the overall shape is reproduced.

Both 1) the total number, and the 2) redshift distribution of the
simulated sources are consistent with the source characteristics of
Abell 1689. Given the results of this comparison, we conclude that the
HDF derived source population distribution is a faithful
representation of the actual source distribution behind massive
lensing clusters.

\subsection{Cluster redshift distribution}

In order to generate a set of massive, strong lensing clusters in
our simulations, we need to derive redshifts from an observed
cluster redshift distribution. To do so, we use the MAssive Cluster
Survey (MACS) \citep{macs2001} sample as our template.  We choose
this sample since it is X-ray luminosity selected, which is a robust
proxy for mass, and the \textit{HST} follow-up of these clusters has
demonstrated that a large majority of these are strong lenses
\citep{macs2005}.

We use the binned MACS cluster redshifts (see fig.~9 in
\citet{macs2001}) to assign $N_{\rm lens}(z)$ to each redshift bin.
The $z\,=\,0.000 - 0.025$ bin is omitted in the analysis since the
cross-section for multiple imaging by clusters in this redshift
range is negligible. An exponential fit is performed and the result
is $N_{\rm lens}(z)\,=\,136.07\exp[-5.845\,z]$. The fit is then
normalized, integrated and inverted to form a probability
distribution. The final probability distribution returns a cluster
lens redshift for use in the simulations in the range $z\,=\,0.025 -
0.600$ and can be expressed analytically as $P_{\rm
lens}(x)\,=\,-0.171\log_{10}[-0.83406(-1.0359+x)] $, where $x$ is a
random number between 0 and 1.

\subsection{Cluster mass distribution}\label{Clusmassgeneration}

\subsubsection{The NFW mass profile}\label{sec:NFW}

The NFW density profile is derived from fits to cluster scale dark
matter haloes that form in \textit{N}-body simulations of a cold
dark matter dominated Universe \citep{nfw1997}. This density profile
is given by,
\begin{equation}\label{eqn:NFWrho}
    \rho(r,z)\,=\,\frac{\rho_0}{((r^2+z^2)^{1/2}/r_{\rm s})(1+(r^2+z^2)^{1/2}/r_{\rm s})^2}
\end{equation}
where $z$ is orientated along the line of sight and the radius $r$
is orthogonal to the line of sight. There are two defining profile
parameters for the NFW model, the characteristic density $\rho_0$
and the scale radius of the profile $r_{\rm s}$. The corresponding
characteristic velocity dispersion $v$ is defined as,
\begin{equation}\label{eqn:NFWvdisp}
    v^2=\frac{4}{3}\,G\,r_{\rm s}^2\rho_0.
\end{equation}
In our simulations we will calculate the total projected mass along
the line of sight within the radius $R\,=\,c\,r_{\rm s}$, where $c$
is the concentration parameter. For $R>r_{\rm s}$ this is given by,
\begin{equation}\label{eqn:NFWtotalmass}
 M(R)=\frac{3\pi v^2 r_{\rm s}}{2 G}\left[\,\ln\left(\frac{R}{2 r_{\rm
 s}}\right)
+\frac{1}{\sqrt{(R/r_{\rm s})^2-1}}\cos^{-1}{\frac{r_{\rm
s}}{R}}\right]
\end{equation}
where $R=c\,r_{\rm s}$ (see \citet{limo2005} for further details on
the profile).

\subsubsection{Masses of the lensing clusters}\label{haloconstruction}

Strong and weak lensing analysis of massive cluster lenses have
determined that the total mass of such clusters within $\sim$ 1 Mpc,
lies in the range, $10^{14}-10^{15}M_\odot$,
\citep{smith2005,lomb2005,pngalgal2006}. This mass range is the
starting point for our construction of simulated smooth cluster
lenses.

Consider a cluster scale halo with an NFW profile. To generate the
halo we first draw a total cluster mass from the range $10^{14} -
10^{15}M_\odot$ uniformly. We then assign profile parameters to
characterize the mass distribution. A concentration is drawn
randomly from the range $c\,=\,4-5$ \citep{bull2001}. Strong lensing
models using NFW fitting profiles find that the typical scale radius
lies in the range $r_{\rm s}\,=\,200 - 400\,$kpc. From the total
mass, $c$ and $r_{\rm s}$, we compute the characteristic velocity
dispersion of the NFW halo. Note that we ignore any scaling of
$\rho_0$ with cosmology for simplicity. Finally, with the halo
structure determined, the ellipticity is then drawn randomly from
$0\leq\epsilon\leq0.15$.

\section{ANALYSIS OF THE SIMULATIONS}\label{sec:analsim}

In this section we discuss the details of our simulations. Firstly,
the procedure used to generate the lensing clusters and multiple
image catalogues are discussed, and we explain how the source
magnitude and redshift distribution and the cluster redshift
distribution are used in the simulations. We then discuss our mass
modeling assumptions. Finally, our marginalization method used to
construct the confidence regions for cosmological parameter recovery
are presented.

\subsection{Cluster and lensed image generation}\label{sec:clusterandimage}

To perform our analysis, we generate clusters that will produce the
requisite number of families of multiple images. The methods used to
generate the cluster lenses are described in detail in
Section~\ref{haloconstruction}. We choose all clusters that produce
multiple images which can be detected via relative astrometry in the
High Resolution Channel (HRC) of ACS. This means the effective
Einstein radius $R_{\rm E}$, which we compute at $z\,=\,6.0$, must
be less than $50\,\arcsec$. In addition to this upper bound we also
apply a lower bound of $R_{\rm E}\,\geq10\,\arcsec$. This bound
restricts the simulations to massive clusters that plausibly exhibit
image configurations that are measurable with the HRC. We note there
is no effect on the simulations if the maximum Einstein radius
condition is dropped, as long as the same pixel resolution is
maintained. This may be the case if observations were taken with the
Wide Field Camera on \textit{HST} and the images were processed with
sub pixel dithering.

Once an appropriate cluster has been generated a source catalogue is
generated. This catalogue is drawn from the HDF distribution derived
in Section~\ref{sec:sourcedistconstruct}. A redshift and magnitude
is assigned for each source using the distribution derived
(numerically) from equation~(\ref{eqn:HDFfit}). Each source is also
assigned a random position in the field of view $4 R_{\rm
E}\,\times\,4 R_{\rm E}$.

The number of sources in the catalogue for which magnitudes,
redshifts and positions are generated, is dictated by the projected
source number density of the ensemble HDF catalogue which is $215$
arcmin$^{-2}$ and the field of view. The source catalogue is then
lensed with the code \textsc{LENSTOOL} \citep{kneib1993,jullo07}.
The lensed catalogue is then sorted into single images and multiple
image families.

At this point we have a complete simulated data set that is
equivalent to a known observed sample of ideal lensing clusters.
Observationally, this can be accomplished efficiently through high
resolution space-based images of known massive clusters or deep
ground based observations of clusters. We note that the MACS cluster
survey provides a catalogue of clusters selected through their X-ray
luminosity. This cluster catalogue provides a useful starting point
for observing programs to identify strongly lensing massive
clusters. Such a systematic survey is currently underway
\citep{macs2005} on \textit{HST} ACS.

\subsection{Image catalogue construction}

The image catalogues we use in our recovery of the cosmological
parameters assume \textit{HST} quality spatial resolution and
spectroscopic redshifts. We first examine the ideal case, where the
image positions and redshifts are known exactly. In Section~8 we
expand our simulations to examine the effects of errors in the image
positions and source redshifts. However, before considering possible
sources of error, a best case scenario must be established when
cluster lenses are stacked.

In the simulations, a magnitude limit is specified  $m_{\rm lim}$,
when constructing the image catalogue from the lensed sources and is
taken to be, unless otherwise stated, $m_{\rm lim}\,=\,24.5$ in the
AB(8140) band \citep{gols2002,broad2005}. This ensures that all
images will have measurable spectroscopic redshifts. Therefore, all
families in the lensed image catalogue which do not have two images
with magnitudes $m\,<\,m_{\rm lim}$ are removed at this point.

Applying the magnitude cut prior to the final selection of families
in the image catalogue is a stringent requirement, since we only use
images with $m\,<\,m_{\rm lim}$ in a given family. In the case of
observational data, although it may not be possible to identify all
members of a given image family spectroscopically, it is often
possible to use properties like the image parity, morphology,
photometric redshifts, or a combination of these to make the
identification. Finally, since core images have been detected in
lensing studies \citep{winn,broad2005}, we do not remove such images
unless they do not pass the magnitude cut.

\subsection{Mass modelling}

In the spirit of our simulations, where we are investigating the
ideal case of stacking strong lensing clusters, we use the same form
of the mass profile that was used to generate the image catalogue.
The centre of the lens is fixed at the origin and the ellipticity
and lens redshift are taken as known from the original cluster lens.
In Section~\ref{sec:lensdep}, we describe how the profile
characteristics of the mass modelling influence the cosmology
recovery. It is shown in Section~\ref{sec:lensdep} that for the NFW
profile, the velocity dispersion and scale radius should be
considered unknown in the cosmology recovery.

We note that the prior of a known mass profile is not the case when
using observational data, where the underlying mass model is
unknown. A variety of mass model priors will need to be considered
in practise, since the mass distribution can be complicated, see for
example the Bullet Cluster of \citet{Bradac2006}. When observational
data is analysed the cosmological parameter bias from the modelling
of realistic clusters must be understood. These additional aspects
present significant computational challenges when stacking a large
number of clusters and we defer their study for the time being.

\subsection{Marginalization and cosmology recovery}

The $\chi^2$ minimization we use is computed in the source plane for
computational speed, since the lens equation is many to $1$ when
mapping from the image plane to the source plane. Our measure is
given by,
\begin{equation}\label{eqn:chi2}
\chi^2= \sum_{i=1}^f \sum_{j=1}^{n_i}
\frac{\|\vec{\beta}_{ij}-\langle\vec{\beta}\rangle_{i}\|^2}{\langle\delta\beta^2\rangle_{ij}}
\label{chi2}
\end{equation}
where $\vec{\beta}_{ij}$ is the source position corresponding to
image $j$ from the $i^{\rm th}$ family as mapped back through the
lens equation (equation~[\ref{lensequation}]), and
$\langle\delta\beta^2\rangle_{ij}$ is the standard deviation squared
in the source plane. The mean source position,
$\langle\vec{\beta}\rangle_{i}$, is defined as
$\langle\vec{\beta}\rangle_{i}=\sum_{j=1}^{n_i}\vec{\beta}_{ij}/n_{i}$
and is the average over all source positions of images belonging to
the $i^{\rm th}$ family. Therefore
$(\vec{\beta}_{ij}-\langle\vec{\beta}\rangle_{i})$ is the difference
between the observed image positions as viewed from the source
plane.

In the calculation of equation~(\ref{eqn:chi2}),
$\langle\delta\beta^2\rangle_{ij}$ is computed using the assumed
error in the image plane of  $\delta\theta=0.1\arcsec$. Using the
known image position, we sample at eight image error points:
$(\theta_x,\theta_y\pm\delta\theta)$,
$(\theta_x\pm\delta\theta,\theta_y)$,
$(\theta_x\pm\delta\theta/\sqrt{2},\theta_y\pm\delta\theta/\sqrt{2})$
and
$(\theta_x\mp\delta\theta/\sqrt{2},\theta_y\pm\delta\theta/\sqrt{2})$,
and map each point back to the source plane using the lens equation.
We then calculate the difference between $\vec{\beta}_{ij}$ and each
of the mapped error points. The eight $\delta\beta$ are first
squared before averaging, i.e. $\langle\delta\beta^2\rangle_{ij}$,
because this over estimates the standard deviation squared relative
to the reverse order. Therefore using this averaging method leads to
a \textit{conservative estimate} of the final $\chi^2$ and thus the
confidence intervals (CIs).

The minimization routine, which uses a parabolic optimization in
combination with parameter space bounds, is designed to ensure the
$\chi^2$ surface of each cluster is accurate to at least
$\delta\chi^2\lesssim0.002$. This accuracy is sought because we want
the final likelihood accurate to $\lesssim1$ per cent. After this is
achieved we form the 1$\sigma$ to 4$\sigma$ CIs.

\section{RESULTS: RECOVERY OF INPUT COSMOLOGY}\label{sec:recovery}

\begin{figure*}
\includegraphics[scale=1]{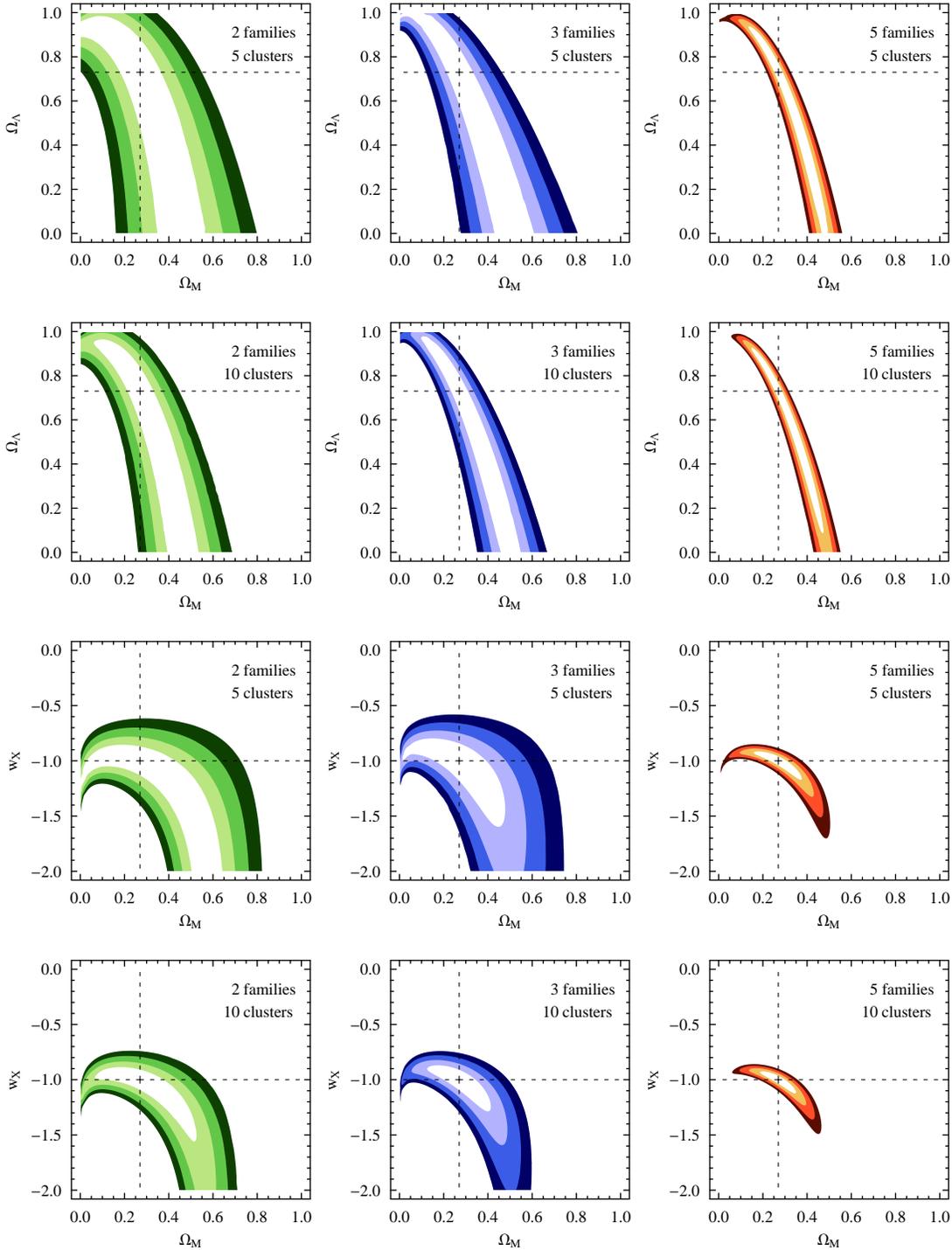}
\caption{Confidence contours on the $\Omega_{\rm
M}$-$\Omega_{\Lambda}$ and $\Omega_{\rm M}$-$\textrm{w}_{\rm X}$
planes for 2, 3 and 5 multiple image families (columns 1, 2 and 3
respectively). For each family number specification, 5 or 10
clusters are stacked. The input cosmology marked by the cross in all
panels, is a flat $\Lambda$CDM universe with $(\Omega_{\rm
M},\Omega_{\rm X},\textrm{w}_{\rm X}) = (0.27,0.73,-1.00)$. In the
top two rows $\textrm{w}_{\rm X}=-1.00$ and a flat universe is not
assumed during the cosmology recovery. In the bottom two rows a flat
cosmology is assumed with $\textrm{w}_{\rm X}$ and $\Omega_{\rm M}$
used in the cosmology recovery. In all panels, the recovery is
undertaken with both the velocity dispersion and the scale radius
allowed to vary. The 1$\sigma$ confidence region is the internal
white region centred on the input cosmology. The 2$\sigma$,
3$\sigma$ and 4$\sigma$ confidence regions are progressively
darker.} \label{fig5and10clusters}
\end{figure*}

\begin{figure*}
\includegraphics[scale=1]{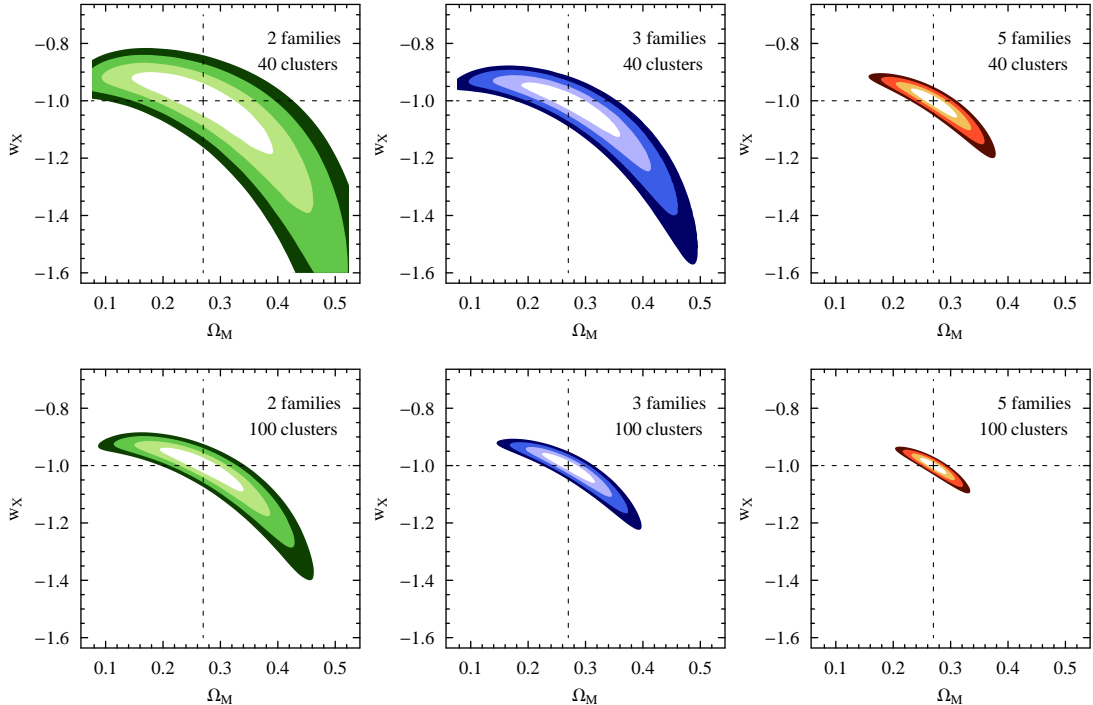}
\caption{Cosmological constraints on the $\Omega_{\rm
M}$-$\textrm{w}_{\rm X}$ plane for 40 and 100 stacked clusters (row
1 and 2 respectively) with 2, 3 and 5 multiple image families
(columns 1, 2 and 3). A flat universe is assumed. Confidence region
shading same as in Fig.~\ref{fig5and10clusters}.}
\label{fig40and100clusters}
\end{figure*}

\begin{table}
  \centering\caption{Recovery Properties on the $\Omega_{\rm M}$--$\textrm{w}_{\rm X}$ plane. The bounds are the limits of the $1\sigma$ confidence
contours, with the input cosmology $(\Omega_{\rm M},\textrm{w}_{\rm
X})=(0.27,-1.00)$ subtracted. $A_1$ is the area enclosed by the
$1\sigma$ contour (see
equation~[\ref{eqn:areaenclosed}]).}\label{tab:simproperties}
\begin{tabular}{cccccc}
\hline \hline Families & Clusters& $\Omega_{\rm M}$ Bounds &
$\textrm{w}_{\rm X}$ Bounds & $A_1$\\\hline
 2& 5  &(-0.27,0.37)&(-1.00,0.15)&0.1825\\
  & 10 &(-0.20,0.26)&(-0.56,0.11)&0.0855\\
  &40  &(-0.10,0.14)&(-0.21,0.08)&0.0227\\
  &100 &(-0.07,0.08)&(-0.10,0.05)&0.0061\\\hline
 3& 5  &(-0.26,0.21)&(-0.60,0.20)&0.1660\\
  & 10 &(-0.14,0.14)&(-0.29,0.13)&0.0447\\
  &40  &(-0.09,0.08)&(-0.12,0.07)&0.0085\\
  &100 &(-0.05,0.05)&(-0.06,0.04)&0.0025\\\hline
 5& 5  &(-0.11,0.11)&(-0.17,0.08)&0.0149\\
  & 10 &(-0.08,0.08)&(-0.12,0.08)&0.0094\\
  &40  &(-0.04,0.04)&(-0.06,0.04)&0.0023\\
  &100 &(-0.02,0.02)&(-0.02,0.02)&0.0004\\\hline
\end{tabular}
\end{table}

We investigate the cosmology recovery properties by considering
cluster samples with different family numbers. To obtain the
recovery properties, we proceed as outlined in Sections
\ref{simulations} and \ref{sec:analsim}. At this stage the input
parameters to the simulations are the number of clusters to stack,
the total number of families in each cluster and the magnitude
limits for the multiple images. The recovery properties are computed
for each cluster, in the specified cosmological parameter space. The
constraints from each cluster in the sample are then stacked to
obtain the final recovery properties.

We consider observations of 5 and 10 clusters, with 2, 3 and 5
multiple image families. The magnitude limit, $m_{\rm lim}$, is
$AB(8140) = 24.5, 24.5$ and $26.0$ for the 2, 3 and 5 multiple image
families respectively and the recovery properties of the input
cosmology are computed in the $\Omega_{\rm M}$-$\Omega_{\rm
\Lambda}$ and $\Omega_{\rm M}$-$\textrm{w}_{\rm X}$ planes. The
results of stacking the first 5 generated clusters and then all 10
are shown in Fig.~\ref{fig5and10clusters} and characterized in
Table~\ref{tab:simproperties} for the $\Omega_{\rm
M}$-$\textrm{w}_{\rm X}$ plane.

In the $\Omega_{\rm M}$-$\Omega_{\rm \Lambda}$ plane we observe that
the parameter constraints obtained by \emph{stacking} lensing
clusters is insensitive to the value of the cosmological constant
when marginalizing over $\Omega_{\rm M}$. The $\Omega_{\rm M}$
constraints are also not stringent, with $\Omega_{\rm M}$
constrained to lie between approximately 0.1 and 0.50 for the 5
family 10 cluster case.

These properties of the cosmology recovery in the $\Omega_{\rm
M}$-$\Omega_{\rm \Lambda}$ are expected. By examining
Fig.~\ref{fig:familyratio}, one can see that the contours passing
through the input cosmology are degenerate for different source
redshift configurations. This means that the constraints derived
from the family ratio will be tightest around the contour defined by
the input cosmology.

However, in the $\Omega_{\rm M}$-$\textrm{w}_{\rm X}$ plane (lower
two rows of Fig.~\ref{fig5and10clusters}) the family ratio is
sensitive to the dark energy equation of state. In
Fig.~\ref{fig:familyratio}, we see that the $\Xi$ contours passing
through the input cosmology are quasi-orthogonal for different
source redshift configurations.  Therefore clusters with distinct
source redshift planes will have different constraint directions in
the $\Omega_{\rm M}$-$\textrm{w}_{\rm X}$ plane. This is also the
case for different lens redshifts given the same two source
redshifts however the effect less pronounced. When \emph{stacking
the cluster sample} the orthogonality of different family ratios
leads to \emph{closed} contours in the $\Omega_{\rm
M}$-$\textrm{w}_{\rm X}$ plane, for the 5 family 10 cluster
simulation.

We then examine the constraints obtained when 40 and 100 clusters
are stacked for 2, 3 and 5 multiple image families. The magnitude
limits are the same as for the 5 and 10 cluster runs but only the
$\Omega_{\rm M}$-$\textrm{w}_{\rm X}$ plane is considered.  The
results are shown in Fig.~\ref{fig40and100clusters} and
Table~\ref{tab:simproperties}. Note that 40 clusters with 5 families
and 100 clusters with only 3 families, produce almost identical CIs.

\section{DEPENDENCE ON LENS PARAMETERS}\label{sec:lensdep}

The role of the lens parameters in the optimal retrieval of the
input dark energy equation of state is examined in this section. We
address the issue of the $\chi^2$ dependence on the marginalization
over combinations of the dominant lens parameters. The
marginalization for each parameter set is computed on the
$\Omega_{\rm M}$-$\textrm{w}_{\rm X}$ plane in a flat Universe, for
an input cosmology $(\Omega_{\rm M},\textrm{w}_{\rm
X})\,=\,(0.27,-1.00)$. We also consider the constraints derived from
a full multi-dimensional marginalization and compare to the
corresponding simple $\chi^2$ minimization.

\subsection{Marginalization over lens
parameters}\label{sec:marglens}

A massive NFW halo is considered in this section and substructure is
omitted from the computations for the purpose of simplicity. The NFW
halo is fixed at the origin and the parametrization we use is
typical of massive cluster lenses \citep{pngalgal2006} and the lens
parameters we use are: $v\,=\,1600\,$km s$^{-1}$, $r_{\rm
s}\,=\,200\,$kpc and $\epsilon=0.10$, with the lens redshift $z_{\rm
l}=0.3$. Two sources are lensed and they are located at
$(x_1,y_1)\,=\,(-1.00,0.10)$ and $(x_2,y_2)\,=\,(1.20,4.80)$
respectively, with redshifts $z_{\rm s1}=1.0$ and $z_{\rm s2}=3.0$.
Sources 1 and 2 have multiplicities of 5 and 3 respectively.

\subsubsection{Case with known lens parameters}

First we consider the case where one has complete knowledge of the
gravitational lens, i.e. all lens parameters are known exactly. The
result of the marginalization with all lens parameters fixed is
shown in the upper left panel of Fig.~\ref{fig:NFWArray} for the NFW
lensing cluster. Note that the confidence region for the NFW profile
is non-zero. In fact, the existence of a non-zero confidence region
is a general feature of the lensing systems in our simulations.

Based on Fig.~\ref{fig:NFWArray} it is clear that the intrinsic
confidence region of a lensing system represents a fundamental limit
on the determination of the cosmological parameters from a single
configuration. Even in the optimal scenario, i.e. complete knowledge
of the lens system, one must consider the confidence region and not
the best fitting cosmology in a rigorous statistical treatment. The
best-fitting cosmological parameters, as for example employed in
\cite{daha2005}, cannot be used to characterize the cosmological
parameter recovery, since they do not take into account the
limitation of the intrinsic confidence region. Noting this property,
we have used CIs exclusively in this work.

\subsubsection{Mass profile parameters}

\begin{figure}
\includegraphics[scale=1]{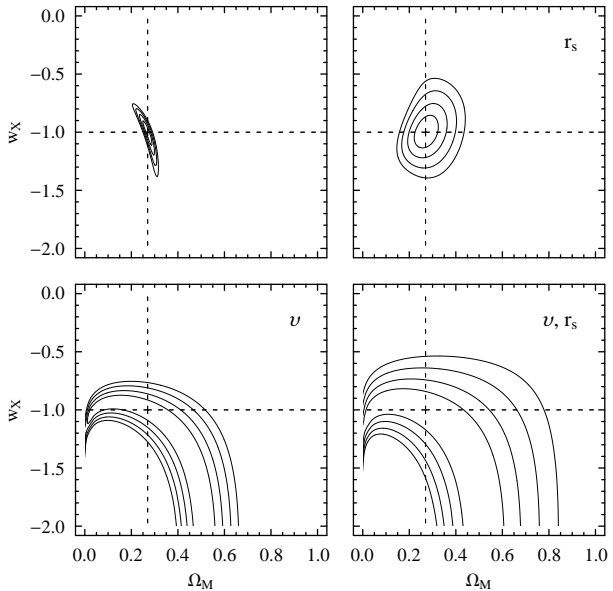}
\caption{Marginalization contours over different NFW parameters. The
1$\sigma$ to 4$\sigma$ contours in the $\Omega_{\rm
M}$-$\textrm{w}_{\rm X}$ plane are shown, and the cross marks the
input cosmology $(\Omega_{\rm M},\textrm{w}_{\rm X}) =
(0.27,-1.00)$. The parameter(s) in the upper right corner of each
panel specify the marginalization variable(s).} \label{fig:NFWArray}
\end{figure}

In practice, one does not have prior information on all the
parameters defining the mass profile of an observed lensing cluster.
For the NFW lensing profile, there are three combinations of the
lens parameters to consider marginalizing over when recovering the
cosmology. These are: the velocity dispersion $v$, the scale radius
$r_{\rm s}$ and both $v$ and $r_{\rm s}$. The results of each
marginalization are given in Fig.~\ref{fig:NFWArray}.

For the NFW profile, the shape of the confidence contours changes
for the different marginalization parameters. For the case of
marginalization over $r_{\rm s}$ alone, in which $v$ is held fixed,
closed confidence contours are obtained in the $\Omega_{\rm
M}$-$\textrm{w}_{\rm X}$ plane. This is distinct from the contours
derived for the marginalization over $v$ alone, which yields a
confidence region which follows the contours of constant
cosmological family ratio $\Xi$. In Fig.~\ref{fig:NFWArray}, the
simultaneous marginalization of $r_{\rm s}$ and $v$ expands the
4$\sigma$ confidence region asymmetrically. In addition, note that
all other confidence contours qualitatively follow the contours for
the family ratio. Therefore it is obvious that optimization over
both the characteristic density and scale radius is required when
recovering cosmology constraints from the NFW profile.

An additional comment on Fig.~\ref{fig:NFWArray} is warranted with
regard to the cosmology recovery. It is obvious from the panels in
Fig.~\ref{fig:NFWArray} that when only $r_s$ or only $v$ are allowed
to vary, that tighter cosmological constraints are obtained,
compared to the case where both $r_s$ and $v$ are marginalized over
simultaneously (bottom right panel of Fig.~\ref{fig:NFWArray}). It
is therefore natural to ask, what are the cosmology recovery
properties in the cases where only $r_s$ or $v$ are marginalized for
stacked clusters? However, when considering actual cluster lenses,
if we were to marginalize over say only $r_s$, we would have to fix
the velocity dispersion. This would require an additional prior, and
in actual cluster lenses this prior is not typically available. It may be
possible to constrain the total cluster mass through X-ray
observations, and in this case, one would use
equation~(\ref{eqn:NFWtotalmass}) to apply the priors to \emph{both}
the velocity dispersion and the scale radius. However, in general,
when using an NFW profile to model an actual strong lensing cluster,
both $r_s$ and $v$ are not known a priori \citep{Comerford2006}. The
reader should therefore note that although stronger cosmological
constraints could be obtained if $r_s$ or $v$ were known a priori,
this is not the case in actual cluster models because separate
priors on $r_s$ or $v$ cannot be found. To this end, we have only
presented the marginalization over both $r_s$ and $v$ in
Section~\ref{sec:recovery}.

\subsection{Full marginalization versus $\chi^2$ minimization}

Up to this point, we have considered a simple minimization routine
of the lens parameters with respect to the $\chi^2$ estimator. We
have done this over the cosmological plane and for different lens
parameter combinations for the NFW profile. However, we now examine
the validity of this assumption for the NFW lensing configuration in
Section~\ref{sec:marglens}.

We now perform a full multi-dimensional marginalization over the
lens parameters, $r_{\rm s}$ and $v$, for the NFW lensing
configuration. Our adopted method is as follows: first, we find the
optimized lens parameters which gives the minimum $\chi_{\rm min}^2$
at each cosmology point. From the optimized lens parameters, we then
compute a $\chi^2$ surface extending to $\chi^2\,=\,(\chi_{\rm
min}^2+50)$. The parameter spacing on the $\chi^2$ surface is
$0.2$\,arcsec for  $r_{\rm s}$ and $0.2\,$km s$^{-1}$ for $v$. The
$\chi^2$ surface is then converted into a likelihood surface using,
$\mathcal{L}\,=\,\exp(-\chi^2/2),$ which is integrated to obtain the
marginalized likelihood at a given cosmology input. This likelihood
is then converted to a $\chi^2$ surface, using $\chi^2\,=\,-2
\ln(\mathcal{L}/\mathcal{L}_{\rm max})$.

To quantify the difference between the multi-dimensional
marginalization and the simple optimization method, we consider an
integral which characterizes the enclosed area,
\begin{equation}\label{eqn:areaenclosed}
    A_n=\int_{R(n)}d\Omega_{\rm M} d\textrm{w}_{\rm X}
\end{equation}
where $n$ defines the confidence contour, and $R(n)$ the region
enclosed by the given contour.

For the multi-dimensional marginalization over $r_{\rm s}$ and $v$,
we have approximately, $A_1 =0.267$, $A_2 = 0.462$, $A_3=0.680$ and
$A_4 =0.889$. Whereas for the simple $\chi^2$ minimization $A_1
=0.273$, $A_2 = 0.464$, $A_3 =0.683$ and $A_4 =0.890$. The
difference between the enclosed area is largest for $A_1$, which is
$\sim2$ per cent. Importantly, the 1$\sigma$ contour from the simple
optimization method encloses an area larger than the full
multi-dimensional marginalization. Both methods recover a $\chi^2 =
0$ at the input cosmology.

Given the maximum $\sim\,2$ per cent difference between the enclosed
areas from the multi-dimensional marginalization and the simple
$\chi^2$-optimization, we are justified in using the simple
optimization method. With sufficient multiple images to constrain
the cluster potential, one expects only one global minimum in the
$\chi^2$ optimization surface. This is the case for the NFW lensing
configuration considered here, and for similar lensing
configurations one expects the optimization method to accurately
reproduce the confidence contours derived from a full
multi-dimensional marginalization.

\section{STUDY OF SURVEY CHARACTERISTICS}

We have also simulated smooth cluster lenses that contain
observational noise. The noise sources we have considered are
positional errors and redshift error. These two sources are the two
major observational parameters which will determine the type of
survey required for the application of CSL. Our simulations are
restricted to the 5 families 10 cluster case (see bottom right panel
in Fig.~\ref{fig5and10clusters}) in the $\Omega_{\rm
M}$-$\textrm{w}_{\rm X}$ plane.

Our objective is to examine the levels of noise which do not lead to
relaxation of the CIs, when compared to our simulations of the ideal
case. We use this approach since this will represent a tolerance
threshold where parameter constraints will weaken.

First we consider redshift accuracy. We have re-run the 5 families
10 cluster simulation with gaussian redshift noise with the
following standard deviations $\Delta z=0.003,0.005,0.01,$ and
$0.1$. When adding the noise to the source redshift, we draw from
the appropriate gaussian and add the resultant value to the known
redshift. Note that all multiple images of a given family receive
the same noise contribution to their redshift. Our results are as
follows. We find that photometric redshift accuracy, i.e. $\Delta
z=0.1$, leads to significant relaxing of the CIs and in some cases
does not recover the input cosmology correctly. Of the remaining
standard deviation simulations, all recover the input cosmology with
increasingly tighter constraints. For $\Delta z=0.003$, there is
negligible confidence contour relaxation when compared to the ideal
simulation.

Next we move onto image positions. Following a similar procedure to
the redshifts, we find that a positional accuracy of $\pm 2$ pixels
of the {\it HST} ACS is required to recover the confidence region in
the $\Omega_{\rm M}$-$\textrm{w}_{\rm X}$ plane. This means that
space based imaging of the strong lensing clusters will be required.
We have also considered image position error in combination with
redshift error, and find that the combination $\Delta z=0.003$ and
$\pm 2$ pixels in ACS does not lead to relaxation of the confidence
contours compared to the ideal case. This means that space based
imaging of strong cluster lenses, followed up by spectroscopic
redshift determination of the multiple images, is the appropriate
survey method to pursue for the application of CSL to cosmography.

\section{LIMITATIONS OF OUR CURRENT ANALYSIS}

As with any other method used to constrain cosmological parameters,
CSL has a number of sources of error one must consider. Our goal has
been to examine observational strategies under the most optimal
conditions in this first detailed study.
Our results demonstrate that by stacking lensing clusters with
several multiple image families, confidence regions for the input
cosmological parameters can be recovered, as shown in
Figs~\ref{fig5and10clusters} and \ref{fig40and100clusters}.

In earlier work, \citet{daha2005} raised the issue of parameter bias
introduced by the unknown cluster mass profile, and the presence of
substructure. They concluded from extracting the best-fitting
cosmological parameters (rather than confidence regions) that this
technique produces $> 100$ per cent errors for a single cluster. We
are not disputing the importance of correctly modelling the mass
profile and substructure. However, in our simulations we have
focused on the recovery properties of the ideal scenario, and the
errors cosmography CSL can tolerate in an observational data set.

\citet{daha2005} has also investigated the recovery bias including
line of sight density fluctuations using a single numerical cluster.
It should also be pointed out that any parameter recovery bias
resulting from substructure along the line of sight (fig. 4 of
\citet{daha2005}), will be minimized if uncorrelated lines of sight
are used when constructing the cosmological constraints. Also, bias
due to line of sight substructure could possibly be accounted for,
by performing redshift surveys behind cluster lenses and modeling
any detected structure along the line of sight in multiple lens
planes. Removing the systematic parameter bias due to substructure
along the line of sight is an important topic that requires further
investigation and is left for future work.

Since we have only used positional accuracy in our cosmological
recovery, we should ask: What additional $\chi^2$ measures can we
add to our current recovery measure of equation (\ref{eqn:chi2})
that will further constrain the input cosmology? The first obvious
candidate is a higher order shape estimators for multiple images,
since we have ignored all shapes and simply used a point
approximation in our simulations. In real clusters multiply imaged
sources will have their shapes distorted as they are strongly lensed
and this will give additional information about the lensing mass
distribution, see for example \citet{Leonard2007}. Another
improvement may come from the possibility of using other
observational maps of clusters to infer additional properties of the
cluster. For example, a total mass constraint can be obtained from
X-ray cluster observations.

\section{CONCLUSIONS}

In this paper, we have examined the recovery of input cosmological
parameters using the cosmological family ratio for several sets of
multiple images in stacked simulated smooth lensing clusters. For a
given lensing configuration, the constraints on the $\Omega_{\rm
M}$-$\textrm{w}_{\rm X}$ plane are tighter than those on the
$\Omega_{\rm M}$-$\Omega_{\Lambda}$ plane. We find that the recovery
of input parameters depends both on the number of multiple image
families with known redshifts and the number of clusters stacked.

As shown in Figs~\ref{fig5and10clusters} and
\ref{fig40and100clusters}, the confidence contours tighten as we
increase the number of families from 2 - 5, and as the total number
of stacked clusters is increased from 1 - 100. For example, the
recovery is comparable in terms of area enclosed within the
1$\sigma$ - 4$\sigma$ contours for the case with 5 families using 40
clusters and 3 families using  100 clusters. As we demonstrate, both
strategies -- stacking fewer clusters with a larger number of image
families or a larger number of clusters with fewer families lead to
comparable constraints.

Currently there are about 40 clusters with high resolution {\it HST}
ACS images, of which a handful (about 7 - 10) have more than 3
families of multiple images with measured redshifts.  Proposed
future space missions, the James Webb Space Telescope and JDEM will
enlarge the existing lensing cluster samples by a factor of 20 or
more, while simultaneously detecting a larger number of families on
average for each cluster through very deep imaging. Determining
spectroscopic redshifts for multiple image families is likely the
limiting step in terms of the observational effort for the CSL
method.

If 10 clusters with 3 multiply imaged sources are stacked, the dark
energy equation of state in a flat universe can be constrained to
$-1.6\lesssim\textrm{w}_{\rm X}\lesssim-0.8$ at 2$\sigma$ where
$\Omega_{\rm M}$ has been marginalized over, see
Fig.~\ref{fig5and10clusters}. This is comparable with current
methods \citep{Komatsu2009}. Therefore to be competitive with
current methods, given that actual massive cluster lenses will
present sub-optimal recovery conditions relative to those considered
here, an initial application of CSL to actual lensing clusters will
require at least 10 clusters with 3 or more multiply imaged sources
with Abell 1689 quality {\it HST} ACS images. Such constraints may
be obtained only if the source redshifts are determined
spectroscopically. These factors indicates that initial observations
should focus on massive clusters that exhibit large lensing cross
sections. Once these strong lensing clusters are imagined to Abell
1689 quality with the {\it HST}, those with 3 or more multiply
imaged sources should be given priority when determining which
lensing clusters to follow up spectroscopically.


\section*{Acknowledgments}

We acknowledge useful discussions with Jean-Paul Kneib.

{}

\bsp

\label{lastpage}

\end{document}